\documentclass[reprint,aps,prb,amsmath,amssymb,superscriptaddress]{revtex4-1}

\usepackage[]{graphicx}
\usepackage[]{units}
\usepackage{times}
\usepackage{bm}
\usepackage[ulem=normalem]{changes}
\usepackage[normalem]{ulem}		% to use \sout command
\usepackage{color}
\usepackage{xr} 				% for cross references
\usepackage{cancel}
\definechangesauthor[color=blue]{SM}

\newcommand{\im}{\Im\textnormal{m}}
\newcommand{\comment}[1]{}

\newcommand\ptwiddle[1]{\mathord{\mathop{#1}\limits^{\scriptscriptstyle(\sim)}}}

\renewcommand{\emph}{\textit}

\begin{document}

\title{Phonon Sidebands in Monolayer Transition Metal Dichalcogenides}
\author{Dominik Christiansen}
\affiliation{Institut f\"ur Theoretische Physik, Nichtlineare Optik und Quantenelektronik, Technische Universit\"at Berlin,  10623 Berlin, Germany}
\author{Malte Selig}
\email{malte.selig@campus.tu-berlin.de}
\affiliation{Institut f\"ur Theoretische Physik, Nichtlineare Optik und Quantenelektronik, Technische Universit\"at Berlin,  10623 Berlin, Germany}
\author{Gunnar Bergh\"auser}
\affiliation{Chalmers University of Technology, Department of Physics, SE-412 96 Gothenburg, Sweden}
\author{Robert Schmidt}
\affiliation{Physikalisches Institut und Zentrum f\"ur Nanotechnologie, Universit\"at M\"unster, 48149 M\"unster, Germany}
\author{Iris Niehues}
\affiliation{Physikalisches Institut und Zentrum f\"ur Nanotechnologie, Universit\"at M\"unster, 48149 M\"unster, Germany}
\author{Robert Schneider}
\affiliation{Physikalisches Institut und Zentrum f\"ur Nanotechnologie, Universit\"at M\"unster, 48149 M\"unster, Germany}
\author{Ashish Arora}
\affiliation{Physikalisches Institut und Zentrum f\"ur Nanotechnologie, Universit\"at M\"unster, 48149 M\"unster, Germany}
\author{Steffen Michaelis de Vasconcellos}
\affiliation{Physikalisches Institut und Zentrum f\"ur Nanotechnologie, Universit\"at M\"unster, 48149 M\"unster, Germany}
\author{Rudolf Bratschitsch}
\affiliation{Physikalisches Institut und Zentrum f\"ur Nanotechnologie, Universit\"at M\"unster, 48149 M\"unster, Germany}
\author{Ermin Malic}
\affiliation{Chalmers University of Technology, Department of Physics, SE-412 96 Gothenburg, Sweden}
\author{Andreas Knorr}
\affiliation{Institut f\"ur Theoretische Physik, Nichtlineare Optik und Quantenelektronik, Technische Universit\"at Berlin,  10623 Berlin, Germany}

%%%%%%%%%%%%%%%%%%%%%%%%%%%%%%%%%%%%%%%%%%%%%%%%%%%%%%%%%%%%% 
\begin{abstract}
Excitons  dominate the optical properties of monolayer transition metal dichalcogenides (TMDs). Besides optically accessible bright exciton states, TMDs exhibit also a multitude of optically forbidden dark excitons. Here, we show that efficient exciton-phonon scattering couples bright and dark states and gives rise to an asymmetric excitonic line shape. The observed asymmetry can be traced back to phonon-induced sidebands that are accompanied by a polaron red-shift. We present a joint theory-experiment study investigating the microscopic origin of these sidebands in different TMD materials taking into account intra- and intervalley scattering channels opened by optical and acoustic phonons. The gained insights contribute to a better understanding of the optical fingerprint of these technologically promising nanomaterials.
\end{abstract}

\maketitle

%%%%%%%%%%%%%%%%%%%%%%%%%%%%%%%%%%%%%%%%%%%%%%%%%%%%%%%%%%%%%%%%%%%%%%%%%%%%%%%%%
%Einleitung

\begin{figure}[t!]
  \begin{center}
     \includegraphics[width=\linewidth]{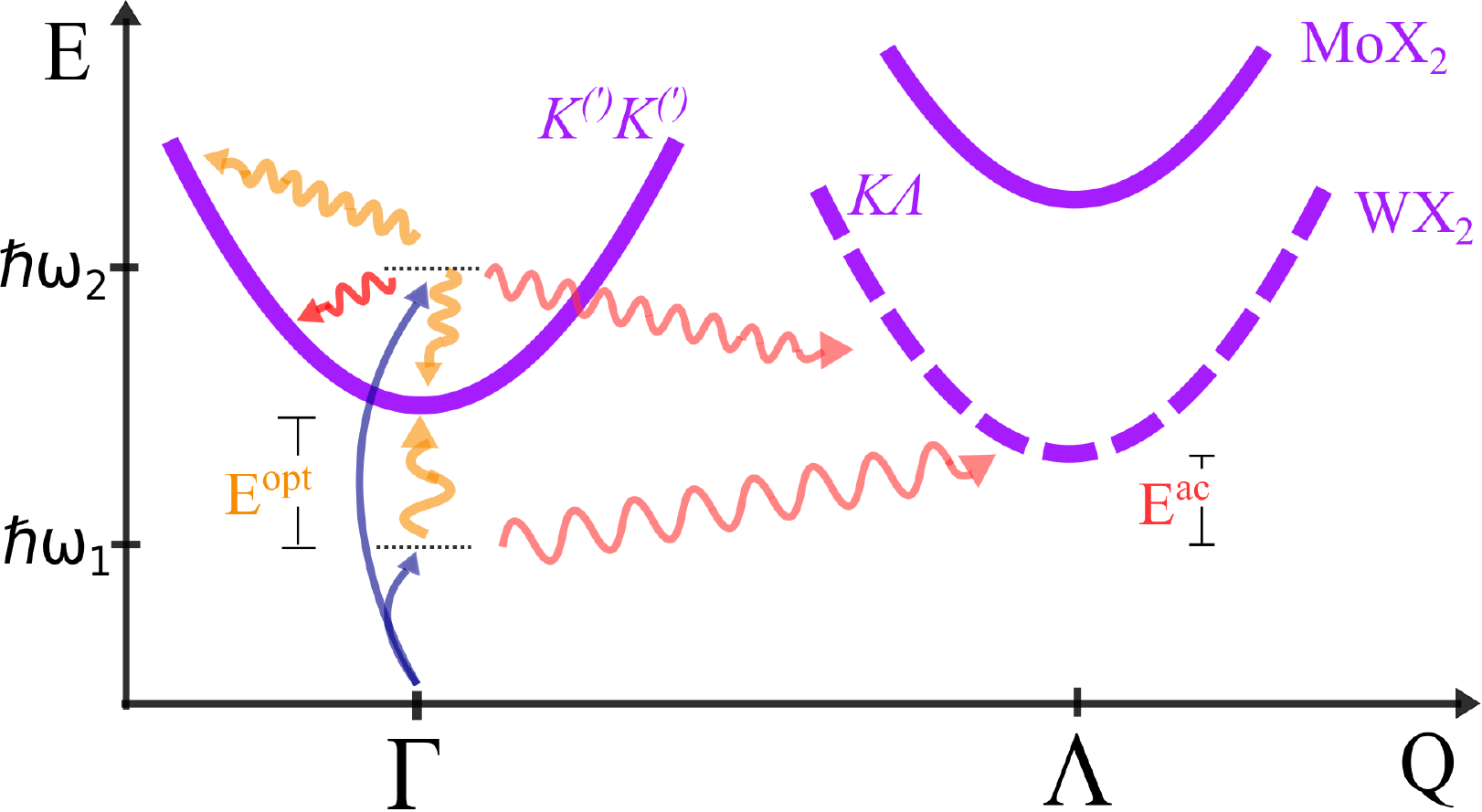}
   \end{center}
    \caption{\textbf{Relaxation channels governing the formation of phonon sidebands.} A radiatively (blue) excited exciton with the energy $\hbar\omega_{1,2}$ at the $\Gamma$ point can scatter through acoustic (red) or optical (yellow) intravalley phonons. In the case of tungsten-based TMDs,  the $K\Lambda$ excitons (dashed line) lie below the $KK$ (or $K'K'$) excitons opening up additional scattering channels through acoustic intervalley phonons. $E^{ac}$ and $E^{opt}$ stand for phonon energies enabling scattering in excitonic states with excitation energies below the dispersion minimum. This quasi-continuous number of scattering channels leads to experimentally accessible phonon sidebands in linear absorption spectra. }
  \label{schema}
\end{figure}

Transition metal dichalcogenides (TMDs) as atomically thin two-dimensional materials show an extraordinarily strong Coulomb interaction leading to the formation of excitons with binding energies of hundreds of meV\citep{He2014,Chernikov2014,Berghauser2014,Ramasub2012,Louie2013}. In combination with a strong light-matter interaction, excitons dominate the optical spectra of these materials. This makes their understanding of crucial importance for TMD-based technology \citep{Butler2013}. Due to a complex quasi-particle band structure\citep{Robert2016,Steinhoff2015,Sanchez2016,Selig2017,Ermin2017}, TMDs possess a variety of dark and bright exciton states\citep{MacDonald2015,Louie2015,Selig2016}. Figure \ref{schema} shows a schematic picture of the excitonic dispersion $E=E(\mathbf{Q})$, where $\Gamma$ point excitons (built up from $K^{(')}K^{(')}$ electronic transitions consisting of electrons in the $K$ ($K'$) valley and holes in the $K$ ($K'$) valley) can be radiatively excited. Subsequently, exciton-phonon scattering within the $\Gamma$ valley and between different excitonic valleys is possible.
In tungsten-based materials, the appearance of dark states below the optical bright state was demonstrated in recent experiments\citep{Arora2015,Zhang2015,Lindlau2017,Lindlau2017_2}. The interaction of excitons with phonons enables scattering of bright excitons from the $\Gamma$ point ($\mathbf{Q=0}$) into optically forbidden dark excitonic states ($\mathbf{Q\neq 0}$ in the $\Gamma$ valley and the excitonic $\Lambda$ valley), which has a strong impact on the excitonic coherence lifetime and on the linewidth of excitonic resonances \citep{Selig2016}.
Recent studies reported phonon-driven homogeneous broadening of these exciton resonances in the range of some tens of meV at room temperature\citep{Moody2015,Dey2016}. Due to the fluctuation-dissipation theorem, beyond line broadening also complex lineshifts including phonon sidebands should exist. Until now, only very weak signatures of phonon sidebands have been observed in bright GaAs quantum films\citep{Kozhernikov1997}. In monolayer TMD materials they have not been investigated yet, in particular nothing is known about the interplay of dark and bright excitons for sideband formation. 

In this work, we present a joint theory-experiment investigation of the exciton-phonon interaction in monolayer TMDs. The theoretical approach is based on excitonic TMD Bloch equations\citep{Robert2016,Selig2016,Thranhardt2000} for the excitonic polarization combined with the Wannier equation\citep{Berghauser2014} providing access to exciton energies and wave functions. Treating the exciton-phonon interaction beyond the Markov approximation, we find polaron red-shifts and strong asymmetric phonon sidebands -- in very good agreement with experimental absorption spectra. 

The goal of the theoretical approach is the calculation of the frequency-dependent absorption coefficient $\alpha(\omega)$ for different TMDs, which is proportional to the imaginary part of the susceptibility. This quantity is determined by the optical matrix element $M_{\mathbf{q}}^{\sigma}$ and the excitonic polarization $P_{\mathbf{Q=0}}^{\mu}(\omega)$. We have introduced the excitonic relative momentum $\mathbf{q}$ and center-of-mass momentum $\mathbf{Q}$ corresponding to the relative and the center-of-mass motion of electrons and holes in the real space. Since  photons have a negligibly small momentum, only excitons with $\mathbf{Q}\approx0$ contribute directly to the absorption. 

To calculate the absorption coefficient (cp. Appendix \ref{matrix}), we need to determine the excitonic wave functions $\varphi_{\mathbf{q}}^{\mu}$ in the state $\mu$ and the expectation value of the microscopic polarization. The wave functions and excitonic eigenvalues $E^{\mu}$, leading to the spectral position of the bright excitonic state, are obtained by solving the Wannier equation\citep{Berghauser2014,Selig2016,Kira2006,Axt2004,Kochbuch}. In this work, we focus on the energetically lowest A 1s  exciton. However, the theory can be expanded in a straightforward way  to higher excitonic transitions. The Coulomb potential is approximated by the Keldysh potential taking explicitly into account a two-dimensional TMD monolayer within a dielectric environment \citep{Berghauser2014,Keldysh1978}. This approach is successful to obtain energies and wavefunctions of TMD excitons\citep{Berghauser2014}, resembling the results from DFT calculations. For the exciton-phonon interaction we included explicitly the exciton-phonon coupling elements for two acoustic and optical modes for intra- and intervalley phonons\footnote{See Appendix for detailed information about the used many-particle Hamiltonian and matrix elements, which includes Ref. [\!\!\citenum{Jin2014,Kormanyos2015,Kaasbjerg2012,Li2013,Kaasbjerg2013}]}.

\begin{figure}[t!]
  \begin{center}
     \includegraphics[width=\linewidth]{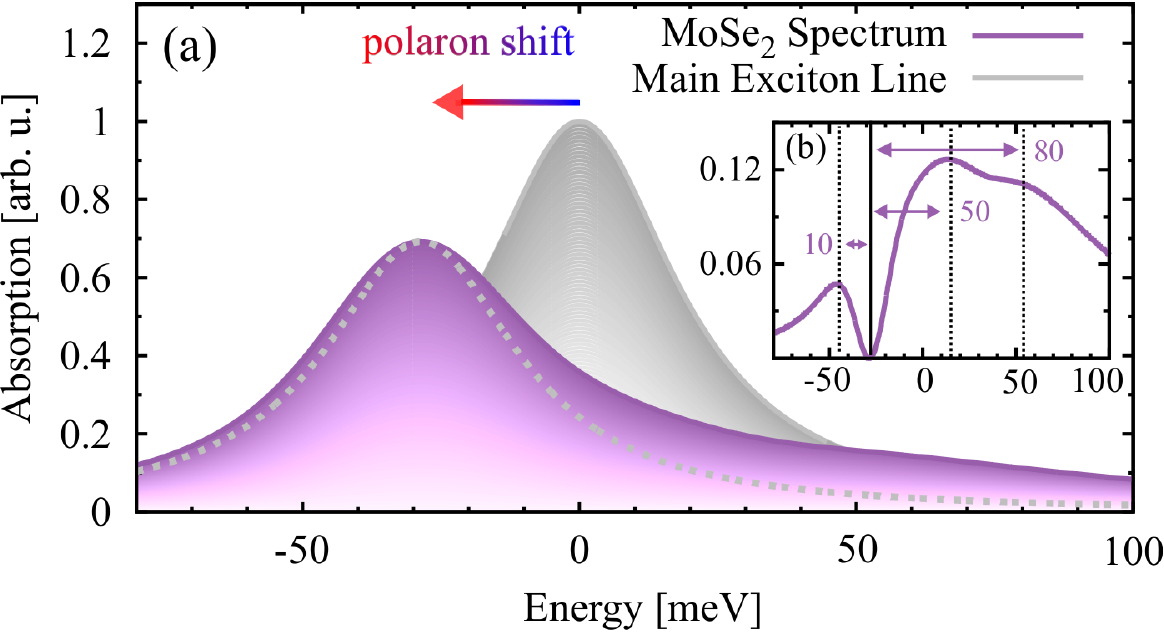}
   \end{center}
    \caption{\textbf{Linear absorption spectrum of monolayer MoSe$_2$ at room temperature showing the presence of phonon sidebands.} (a) The absorption spectrum of MoSe$_2$ as an exemplary TMD material at room temperature illustrates phonon sidebands. The filled grey curve shows the main exciton line revealing a clear polaron red-shift in the presence of phonons. For reasons of clarity, the shifted main exciton line (gray dashed) is also depicted. 
(b) Differential spectrum line showing the positions of the sidebands with respect to the main resonance (black solid line) due to the absorption and emission of phonons.}
  \label{scheme}
\end{figure}

The Heisenberg equation for the excitonic polarization leads to a coupled system of differential equations for the polarization and phonon-assisted polarization solved in a self-consistent second order Born approximation\footnote{See Appendix for details about the Bloch equations, which includes Ref. [\!\!\citenum{Knorr1996,Evgeny2012}]}. A direct solution in frequency domain allows us to consider non-Markovian processes.
We analytically obtain for the absorption spectrum
\begin{align}
\alpha(\omega)=C_0\im\left(\frac{\mid\sum_{\mathbf{q}}M_{\mathbf{q}}^{\sigma}\varphi_{\mathbf{q}}^{\mu}\mid^2}{\hbar\omega-E^{\mu}(\mathbf{0})+i\gamma_{\mathbf{0}}^{1\mu}-\Sigma_{\mathbf{0}}^{\mu}(\omega)}\right) \label{a}
\end{align}
with the self-energy $\Sigma_{\mathbf{Q}}^{\mu}(\omega)$ defined in Eq. \ref{G} and the constant $C_0=\hbar e_0/m_0\epsilon_0\omega nc_0$, where $e_0$, $m_0$, $c_0$, $n$ are the elementary charge, the free electron mass, the speed of light in vacuum and the refractive index, respectively. The optical matrix element and the wave function determine the strength of the light-matter coupling. The excitonic energy and the self-energy determine the polaron shift of the main exciton line and phonon sidebands. The phonon-induced line broadening $\gamma_{\mathbf{0}}^{1\mu}$ is self-consistently included\citep{Schilp1994,Selig2016}.

Figure \ref{scheme} (a) shows the comparison of the absorption spectrum of monolayer MoSe$_2$ at room temperature with and without including the non-Markovian exciton-phonon contribution. We find a polaron red-shift of approximately \unit[29]{meV} with respect to the unshifted exciton main line (gray). Furthermore, the spectrum exhibits asymmetric phonon sidebands, in particular at the high-energy side of the main resonance\citep{Ostreich1997,Krummheuer2002,Forstner2002,*Forstner2003}. Subtracting the homogeneous linewidth shows that we observe multiple sidebands above the main resonance originating from emission and absorption of acoustic and optical phonons, inset: Fig. \ref{scheme}.  Additionally, we also find a phonon sideband located approximately \unit[10]{meV} below the main peak. This peak can be traced back to the absorption of optical phonons during the optical absorption of low energy light. Due to scattering into dark excitonic states with $\mathbf{Q\neq 0}$ along the exciton dispersion $E^{\mu}(\mathbf{Q})$ of the $KK$ exciton (intravalley scattering), we find a continuous absorption up to the resonance energy. Exciton-phonon scattering into the A 2s state is not very likely due to the fact that we excite close to the A 1s exciton and that there is a large energetic difference between A 1s and A 2s excitons\citep{Berghauser2014}.

A detailed study of the MoSe$_2$ spectrum in Fig. \ref{scheme}, consisting of switching on and off the different phonon modes and valleys shows that the main contribution to the linewidth results from intravalley scattering processes in the $\Gamma$ valley. Besides the interaction with acoustic phonons, we find that optical phonons contribute to the scattering at higher energies. The position of the sidebands is at higher energies compared to the corresponding phonon energies, since there is an additional dependence on the exciton-phonon coupling element and frequency: The exact spectral position is determined by the self-energy 
\begin{align}
\Sigma_{\mathbf{Q}}^{\mu}(\omega)&=\sum_{\mathbf{q}',\alpha,\pm}\frac{\mid g_{\mathbf{q}'}^{\mu\alpha}\mid^2\left(\frac{1}{2}\pm\frac{1}{2}+n_{\mathbf{q}'}^{\alpha}\right)}{\hbar\omega-E^{\mu}(\mathbf{Q}+\mathbf{q}')\mp E_{\pm\mathbf{q}'}^{\alpha}+i\gamma_{\mathbf{Q}+\mathbf{q}'}^{2\mu}}. \label{G}
\end{align}
The subscripts $\alpha$ and $\mathbf{q'}$ denotes the phonon mode and momentum. To obtain microscopic insight into the character of the sidebands, we find an analytical expression in a limiting case: First, we simplify Eq. \ref{G} by setting the self-consistently calculated damping constant of the phonon-assisted polarization (shown in the Appendix \ref{inhom}) $\gamma_{\mathbf{Q}+\mathbf{q}'}^{2\mu}$ equal to zero. It contributes to a broadening of the resonance, but has only minor influence on its position. Then, we choose only one intravalley phonon mode, so that the $\alpha$- and $\mathbf{q}'$-sum vanish. Setting the denominator in Eq. \ref{G} zero, we find an expression for the relative distance between the main resonance and the phonon sideband. The position either below the dispersion minimum (-) caused by absorption or emission or above the dispersion minimum (+) due to absorption or emission of an acoustic or optical phonon (cf. Fig. \ref{scheme}) reads $\Delta E_{\mp}=\frac{\mp E^{\alpha}}{2}\mp\sqrt{\frac{(E^{\alpha})^2}{4}+\eta_{\alpha}\mid g^{\mu \alpha}\mid^2}$. Here, $\eta_{\alpha}$ corresponds to the Bose-Einstein distribution $n_{\alpha}$ for absorption and to $(1+n_{\alpha})$ for emission of phonons. The relation shows that the spectral position of phonon sidebands depends on the energy of the involved phonon and the square of the exciton-phonon coupling element times the phonon population. As a result, the stronger the coupling, the larger is the distance between the main peak and the phonon sideband. We apply this analysis to Fig. \ref{scheme} (b): In MoSe$_2$, LO phonons have an energy of \unit[37]{meV} and the coupling element is approximately  \unit[50]{meV}. Inserting these values in the equation, we find a phonon sideband at approximately \unit[10]{meV} below the main resonance stemming from the absorption of LO phonons. Furthermore, we find two sidebands located approximately \unit[50]{meV} and \unit[80]{meV} above the main resonance reflecting absorption and emission of LO phonons, respectively. For acoustic phonons we find due to the linear dispersion relation a continuous broadening close to the main resonance. All of these results are in agreement with the absorption line shape in Fig. \ref{scheme} (a).
The analysis made so far shows that intravalley scattering channels are dominant. The coupling to the broad range of dark $KK$ excitonic states leads to a broadening mainly towards higher energies. Intervalley scattering involving the dark $K\Lambda$ excitonic states is not very efficient in MoSe$_2$, since here the intervalley dark exciton states $K\Lambda$ lie \unit[100]{meV} above the bright exciton (cf. Fig. \ref{schema}) suppressing all phonon-induced intervalley scattering processes, fortifying the asymmetry. We will see below that the situation is different for  WSe$_2$, where phonon-induced intervalley scattering will be important.
\begin{figure}[t!]
  \begin{center}
    \includegraphics[width=\linewidth]{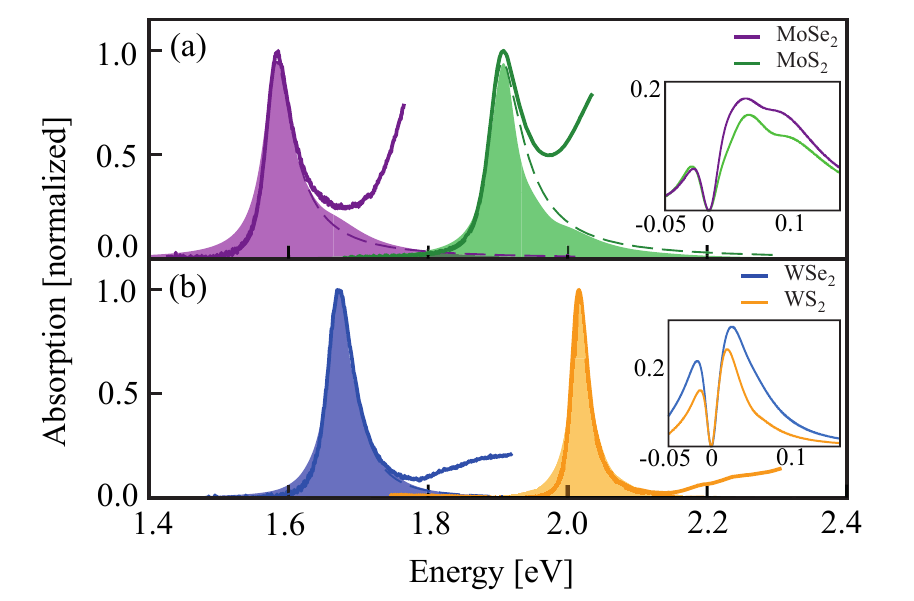}
  \end{center}
  \caption{\textbf{Theory-experiment comparison of the absorption of different TMDs}. Normalized experimental (solid line) and theoretical (filled area) absorption spectra. The contribution of the A exciton in the experimental spectra is indicated by the dashed curves. (a) shows molybdenum-based TMD monolayer spectra and calculated differential spectra as inset, while (b) contains the tungsten-based TMDs. We find less asymmetric spectra for tungsten- than for molybdenum based TMDs. Furthermore, TMDs with sulfur show less broadened phonon sidebands, which can be ascribed to lower phonon population numbers. For comparison, the theoretical curves are shifted on top of the experimental curves.}
 \label{TMDs}
\end{figure}
Besides the discussed momentum-forbidden dark excitonic states, spin-forbidden excitons with Coulomb-bound electrons and holes with opposite spins have been studied in literature\citep{MacDonald2015,Louie2015,Echeverry2016}. While these states can be very important for the efficiency of photoemission, we expect their impact  on the excitonic line shape to be rather weak. The reason is that they require spin-flip processes, which were shown to occur on a picosecond time scale\citep{Glazov2014} that is at least one order of magnitude slower than the considered exciton-phonon scattering channels. Nevertheless, further studies are needed to quantify the contribution of spin-forbidden states.

So far, we have only investigated MoSe$_2$ as an exemplary TMD material. Now, we extend our study to other semiconducting TMDs including MoS$_2$, WS$_2$,  and WSe$_2$ and compare the calculations to experiments. To analyze the experiments we need to take into account that depending on the material, A and B exciton, as well as excited states of the A and B exciton partly overlap. As before, in theory, we consider only the absorption of the A exciton. Therefore, we fit the absorption maxima of the measured spectra and use only the contribution of the A exciton for the comparison. Note that the accuracy of the fit depends on the overlap of the A and B exciton which is mostly significant for MoS$_2$\footnote{See Appendix for the details of our measurement and fitting procedure, which includes Ref. [\!\!\citenum{Kendall}]}. We believe that our work will trigger further coherent nonlinear experiments,
such as two-dimensional spectroscopy, in particular four wave mixing\citep{Jakubczyk2016}, where due to the non-Markovian character of the exciton-phonon interaction  the temporal dynamics of the  emission should deviate from a simple exponential decay expected from symmetric Lorentzian broadened lines.

  Figure \ref{TMDs} displays the calculated (filled) and the experimental absorption (solid line) of the A exciton of different TMDs at room temperature. Since the theory does not provide the full temperature-dependent shift of the exciton line (see comments below), the theoretical curves are shifted on top of the corresponding measured spectra.  Generally, we find a good agreement between the theoretically predicted and experimentally measured spectra. Especially the linewidth of the A exciton is accurately described by the theory. Interestingly, we observe that all experimental peaks are slightly narrower on the low energy tail than in the calculation. Our theory seems to overestimate the coupling to short-range optical phonons. The only theoretical curve showing stronger deviation from the experimental result is MoS$_2$. One reason for the discrepancy might be that we have only included exciton-phonon scattering but neglected broadening by e.g. disorder-induced channels. The MoS$_2$ monolayer is exfoliated from a natural crystal (all other TMDs have been grown artificially) and is therefore expected to have a higher degree of disorder than the other TMD  materials (see Appendix).

We find less asymmetric spectra for tungsten- in comparison to molybdenum-based materials. This can be mainly traced back to different exciton-phonon coupling elements and the different relative energetic position between the  $K\Lambda$ and  $KK$ excitons, cf. Fig. \ref{schema}. In W$X_2$, we find  weaker exciton-phonon coupling elements\citep{Selig2016} accounting for narrower intravalley sidebands compared to Mo$X_2$. First, the small coupling for W$X_2$ excitons to optical phonons\citep{Selig2016} leads to weak optical sidebands and correspondingly, a smaller broadening. Second, the less pronounced sidebands in W$X_2$ can be ascribed to the strongly efficient intervalley scattering between $KK$ and $K\Lambda$ excitonic states, since the dark $K\Lambda$ states lie below the bright $KK$ exciton. Due to the favourable energetic position we find also phonon absorption and emission processes into the $K\Lambda$ excitonic states with excitation energies below the dispersion minimum. Therefore, the optically bright state couples resonantly to a broad band of $K\Lambda$ states. This leads to a broadening of the A exciton resonance at both energetic sides of the main resonance and thus a less pronounced asymmetry. We conclude that the main asymmetry stems from intravalley scattering, while the $K\Lambda$ states contribute mostly to the broadening. In Fig. \ref{TMDs}, we also observe differences depending on the involved chalcogen atom. TMDs involving selenium show a more efficient exciton-phonon scattering compared to TMDs with sulfur. The explanation lies in the smaller phonon energies for Se-TMDs, which results in higher phonon populations at fixed temperature. This has a direct impact on the phonon sidebands as well as the polaron shift.
\begin{figure}[t!]
 \begin{center}
\includegraphics[width=\linewidth]{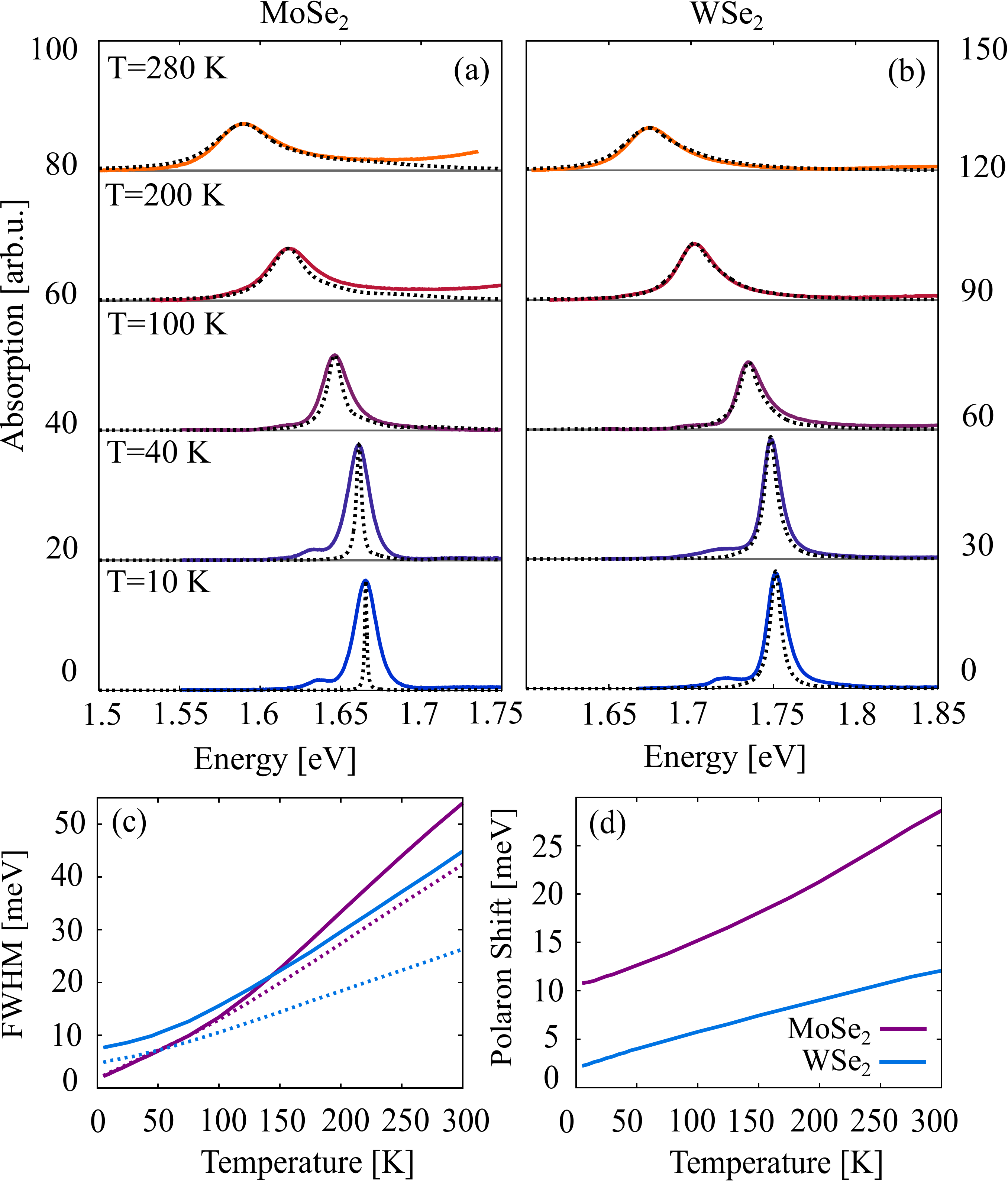}
 \end{center}
 \caption{\textbf{Temperature-dependent absorption spectra.} (a) and (b) Theoretical spectra (dashed) shifted on top of the experimental (continuous) curves for MoSe$_2$ and WSe$_2$. The homogeneous broadening of excitonic resonances becomes larger at increased temperatures. At the same time, the phonon sidebands become more pronounced. This is accompanied by a growing red-shift. (b) The polaron shift shows a non-linear dependence on temperature and is larger for molybendum- than for tungsten-based TMDs. (c) Full width at half maximum (FWHM) of excitonic resonances including phonon sidebands and considering only the main exciton line (dashed) of MoSe$_2$ and WSe$_2$. Due to possible phonon-induced intervalley scattering to the dark $K\Lambda$ excitonic states (cf. Fig. \ref{schema}), WSe$_2$ shows a higher FWHM at low temperatures.}
 \label{spectrum}
\end{figure}

A crucial quantity for exciton-phonon coupling is temperature.  Figure \ref{spectrum} (a) and (b) shows the experimental (solid) and theoretical (dashed) absorption spectra of MoSe$_2$ and WSe$_2$ for temperatures from \unit[10]{K} up to the room temperature. First, we can observe a clearly growing line broadening and red-shift with increasing temperature. Second, as an important qualitative signature, we observe that the asymmetry of the spectrum due to phonon sidebands becomes much more pronounced with increasing temperature. At \unit[10]{K}, the sidebands are almost invisible due to the small phonon population strongly reducing the self-energy. The higher the temperature, the larger is the phonon population, and the more pronounced are the sidebands. Simultaneously, we have a competing line broadening which is growing with the temperature. Furthermore, at low temperatures we see a strong broadening of the experimental spectra due to disorder. However, we see that the asymmetry of the excitonic line shape is a temperature-dependent effect emphasizing the phonon induced origin of the inhomogeneous broadening. Fig. \ref{spectrum} (c) shows a direct comparison between the homogeneous (main exciton line) and phonon sideband broadening (FWHM). The first shows a linear dependence on temperature due to acoustic phonons \cite{Selig2016}. Above approximately \unit[100]{K}, we can observe a deviation from the linear dependence reflecting the formation of phonon sidebands. We find a remarkably different temperature dependence of the FWHM. For  WSe$_2$, the latter is larger than for MoSe$_2$ at small temperatures, however it increases slower, so that the value for MoSe$_2$ takes over at temperatures  above \unit[150]{K} due to the stronger exciton-phonon coupling. The larger width of WSe$_2$ at lower temperatures is due to the fact that the dark  $K\Lambda$ exciton lies energetically lower than the bright $KK$ exciton opening up phonon-induced scattering channels between these states even at very low temperatures. 

Figure \ref{spectrum} (d) shows the calculated polaron red-shift which is smaller than the measured red-shift, cf. Fig. \ref{spectrum} (a) and (b). However, we note that the calculated polaron shift reflects only a part of the temperature-dependent red-shift of optical resonances, which usually is described by the phenomenological Varshni relation\cite{Varshni1967}. To obtain the total change of the band gap, the temperature-dependent lattice expansion, for all resonant and off-resonant phonon modes (higher order contributions beyond second order Born approximation), has to be taken into account. This goes beyond the scope of our work, where the focus lies on the resonantly interacting phonons, since they are responsible for the asymmetric exciton peak shape. We observe a clearly smaller polaron red-shift for WSe$_2$ compared to MoSe$_2$, cf. Fig. \ref{spectrum} (d). This is due to the smaller exciton-phonon coupling\citep{Selig2016}. 

In conclusion, we present a non-Markovian investigation of polaronic effects on the excitonic absorption spectra in monolayer TMDs. We find that phonon-assisted polarizations leave a fingerprint in form of polaron red-shifts and phonon sidebands, which is in excellent agreement with  experimental absorption spectra. The observed frequency-dependent broadening and shifts lead to a deviation from the symmetric Lorentzian line shape. In particular, intravalley phonons of different optical and acoustic modes enable a coupling between bright and dark excitonic states giving rise to a strong asymmetry of the excitonic resonance. Interestingly, the coupling to dark intervalley excitonic states leads to a less pronounced asymmetry. This leads to a qualitative difference in the exciton-phonon coupling for molybdenum- and tungsten-based TMDs. 

We acknowledge financial support from the Deutsche Forschungsgemeinschaft (DFG) through SFB 910 (D.C., A.K.), SFB 787  (G.B.,M.S.,A.K.) and SFB 658 (E.M.). M.S. thanks the school of nanophotonics (SFB 787) for support. This project has also received funding from the European Unions Horizon 2020 research and innovation programme under grant agreement No 696656 (Graphene Flagship, E.M.) and No. 734690 (SONAR, A.K.). A.A. acknowledges the financial support from Alexander von Humboldt foundation and DFG project No. AR 1128/1-1. Finally, we thank Hendrik Paysen (TUB) for valuable discussions.

\appendix

\section{Matrix elements} \label{matrix}
The optical matrix element determines the selection rules and the oscillator strength. In context of the dipole approximation it reads
\begin{align}
\mathbf{M}_{\mathbf{k}\mathbf{k}'}^{\lambda\lambda'}=-\frac{e}{m}\frac{\hbar}{i}\mathbf{A}\langle\Psi^{\lambda}(\mathbf{k},\mathbf{r})\mid\nabla\mid\Psi^{\lambda'}(\mathbf{k}',\mathbf{r})\rangle \label{OME}
\end{align}
and can be calculated analytically within a tight-binding approximation. Because of the optical excitation we can simplify $\lambda\neq\lambda'$ and $\mathbf{k}=\mathbf{k}'$. \newline
The Coulomb matrix element
\begin{align}
V^{a~b}_{c~d}=\langle\Psi_a(\mathbf{r})\Psi_b(\mathbf{r'})\mid V(\mathbf{r-r'})\mid\Psi_d(\mathbf{r'})\Psi_c(\mathbf{r})\rangle
\end{align}
determines the interaction strength for electronic transitions from state $(a,b)$ to state $(c,d)$. Equally, it can be calculated in context of a tight-binding approximation considering in zero-th order of the Coulomb potential intraband interaction and in first order additionally interband interaction. \newline
The exciton-phonon matrix element is defined as a sandwich between the excitonic wavefunctions of initial and final state and the electron-phonon matrix element
\begin{align}
g_{\mathbf{q}'}^{\mu\lambda\alpha}=\sum_{\mathbf{q}}\varphi_{\mathbf{q}}^{*\mu}g_{\mathbf{q},\mathbf{q-q}',\mathbf{q}'}^{c\alpha}\varphi_{\mathbf{q}-\beta\mathbf{q}'}^{\lambda}-\varphi_{\mathbf{q}}^{*\mu}g_{\mathbf{q+q}',\mathbf{q},\mathbf{q}'}^{v\alpha}\varphi_{\mathbf{q}+\alpha\mathbf{q}'}^{\lambda}.
\end{align}
The electron phonon matrix element is treated within a deformation potential approximation using values of DFT calculations \cite{Jin2014}. We included explicitly exciton-phonon matrix elements for the LA, TA, LO and TO mode.

\section{Computation of the inhomogeneous broadened spectrum} \label{inhom}

Using Heisenberg's equation of motion, the TMD Bloch equation for the microscopic polarization in the excitonic basis reads
\begin{align}
\partial_tP_{\mathbf{Q}}^{\mu}(t)&=\frac{1}{i\hbar}\left(E^{\mu}(\mathbf{Q})-i\gamma^{1\mu}_{\mathbf{Q}}\right)P_{\mathbf{Q}}^{\mu}+\sum_{\mathbf{q}}\Omega_{\mathbf{q}}^{cv}\varphi_{\mathbf{q}}^{*\mu}\delta_{\mathbf{Q},0} \nonumber \\\label{PGI}
&+\frac{1}{i\hbar}\sum_{\mathbf{q}',\lambda,\alpha}g^{\mu\lambda\alpha}_{\mathbf{q}'}\left(S_{\mathbf{Q}+\mathbf{q}',\mathbf{q}'}^{\lambda\alpha}+\tilde{S}_{\mathbf{Q}+\mathbf{q}',-\mathbf{q}'}^{\lambda\alpha}\right). 
\end{align}
The first term describes the oscillation of the excitonic polarization with $E^{\mu}(\mathbf{Q})=\frac{\hbar^2\mathbf{Q}^2}{2M}+E^{\mu}$. The excitonic dispersion is determined by the center-of-mass momentum $\mathbf{Q}$. The appearing damping constant $\gamma^{1\mu}_{\mathbf{Q}}$ consists of the phonon-induced homogeneous dephasing of the excitonic polarization \citep{Thranhardt2000} and the radiative coupling, i.e spontaneous recombination of an electron and a hole under emission of a photon. The latter is obtained by self-consistently solving the Bloch equation and the Maxwell equation in a two-dimensional geometry for the vector potential $A^{\sigma_-}$ \citep{Selig2016,Knorr1996}. The second part in Eq. (\eqref{PGI}), coming from the carrier-light Hamiltonian, describes the optical excitation with the Rabi frequency $\Omega_{\mathbf{q}}^{\sigma}=\frac{e}{m}A^{\sigma}M_{\mathbf{q}}^{\sigma}$. We consider right-handed circular polarized light $\sigma=\sigma_-$ for excitation in the K valley \citep{Berghauser2014}. \newline
The third part in Eq. (\eqref{PGI}) stems from the exciton-phonon interaction, where phonon-assisted polarizations $\ptwiddle{S}_{\mathbf{Q},\mathbf{q}'}=\sum_{\mathbf{q}}\varphi_{\mathbf{q}}^{*\mu}\langle a_{\mathbf{q}+\beta\mathbf{Q}}^{\dagger v}a_{\mathbf{q}-\alpha\mathbf{Q}}^{c}b_{\mathbf{q}'}^{(\dagger)}\rangle$ appear, describing scattering driven by phonon absorption and  emission. We again exploit the Heisenberg equation to calculate the time evolution of $S_{\mathbf{Q},\mathbf{q}'}$, $\tilde{S}_{\mathbf{Q},\mathbf{q}'}$ to obtain a closed set of equations. We find in the limit of linear optics
\begin{align}
\partial_tS_{\mathbf{Q}+\mathbf{q}',\mathbf{q}'}^{\mu\alpha}&=\frac{1}{i\hbar}\left(E^{\mu}(\mathbf{Q}+\mathbf{q}')+E_{\mathbf{q}'}^{\alpha}-i\gamma_{\mathbf{Q+q'}}^{2\mu}\right)S_{\mathbf{Q}+\mathbf{q}',\mathbf{q}'}^{\mu\alpha} \nonumber \\
&+\frac{1}{i\hbar}\sum_{\lambda}g_{-\mathbf{q}'}^{\mu\lambda\alpha}(1+n_{\mathbf{q}'}^{\alpha})P_{\mathbf{Q}}^{\lambda} \label{SGl} \\
\partial_t\tilde{S}_{\mathbf{Q}+\mathbf{q}',-\mathbf{q}'}^{\mu\alpha}&=\frac{1}{i\hbar}\left(E^{\mu}(\mathbf{Q}+\mathbf{q}')-E_{-\mathbf{q}'}^{\alpha}-i\gamma^{2\mu}_{\mathbf{Q+q}'}\right)\tilde{S}_{\mathbf{Q}+\mathbf{q}',-\mathbf{q}'}^{\mu\alpha} \nonumber \\
&+\frac{1}{i\hbar}\sum_{\lambda}g_{-\mathbf{q}'}^{\mu\lambda\alpha,s}n_{\mathbf{q}'}^{\alpha}P_{\mathbf{Q}}^{\lambda}. \label{SGl2}
\end{align}
The first term in the equations describes the free oscillation of the phonon-assisted polarizations with the excitonic eigenenergy $E^{\mu}$, the phonon energy $E_{(-)\mathbf{q}'}^{\alpha}$, and the kinetic energy $\frac{\hbar^2(\mathbf{Q}+\mathbf{q}')^2}{2M}$. The damping constant $\gamma_{\mathbf{Q+q'}}^{2\mu}$ stands for the phonon-induced dephasing of the excitonic polarization in a self-consistent Born approximation. The last term expresses the exciton-phonon coupling with the matrix element $g_{\mathbf{q}'}^{\mu\lambda\alpha}=\sum_{\mathbf{q}}\varphi_{\mathbf{q}}^{*\mu}g_{\mathbf{q},\mathbf{q-q}',\mathbf{q}'}^{c\alpha}\varphi_{\mathbf{q}-\beta\mathbf{q}'}^{\lambda}-\varphi_{\mathbf{q}}^{*\mu}g_{\mathbf{q+q}',\mathbf{q},\mathbf{q}'}^{v\alpha}\varphi_{\mathbf{q}+\alpha\mathbf{q}'}^{\lambda}$ that is built by the electron-phonon matrix element sandwiched by excitonic wave functions respecting the excitonic phase. The phonon population $n_{\mathbf{q}'}^{\alpha}$ follows the Bose-Einstein statistics. We solve the coupled system of differential equations \eqref{PGI}-\eqref{SGl2} by Fourier transformation considering non-Markovian processes \cite{Evgeny2012}. Inserting the phonon-assisted polarizations in Eq. (\eqref{PGI}) we obtain
\begin{align}
P_{\mathbf{Q}}^{\mu}(\omega)=\frac{i\hbar\sum_{\mathbf{q}}\Omega_{\mathbf{q}}^{cv}\varphi_{\mathbf{q}}^{*\mu}\delta_{\mathbf{Q},0}}{\hbar\omega-E^{\mu}(\mathbf{Q})+i\gamma^{1\mu}_{\mathbf{Q}}-\Sigma_{\mathbf{Q}}^{\mu}(\omega)} \label{eq:P}
\end{align}
with the self-energy $\Sigma_{\mathbf{Q}}^{\mu}(\omega)$:
\begin{align}
\Sigma_{\mathbf{Q}}^{\mu}(\omega)&=\sum_{\mathbf{q}',\alpha,\pm}\frac{\mid g_{\mathbf{q}'}^{\mu\alpha}\mid^2\left(\frac{1}{2}\pm\frac{1}{2}+n_{\mathbf{q}'}^{\alpha}\right)}{\hbar\omega-E^{\mu}(\mathbf{Q}+\mathbf{q}')\mp E_{\pm\mathbf{q}'}^{\alpha}+i\gamma_{\mathbf{Q}+\mathbf{q}'}^{2\mu}}. \label{selbstE}
\end{align}
Inserting the expression for $P_{\mathbf{Q=0}}^{\mu}(\omega)$ in Eq. \eqref{a}, we find the absorption spectrum under influences of phonon-assisted polarizations. 

\section{Optical transmission spectroscopy on TMD monolayers} \label{experiment}

The samples are prepared by mechanical exfoliation from single crystals onto borosilicate glass with a thickness of \unit[170]{$\mu$m}. The transmission measurements are performed using the broadband emission from a tungsten halogen lamp for the selenide based materials and a white light emitting diode for the sulfides. The white light illuminates the sample in a wide-field configuration through a microscope objective. A second microscope objective is used to collect the transmitted light from an area of less than  $\unit[1]{\mu m^2}$ and to image the sample plane onto a charged coupled device attached to an imaging spectrograph ($f$=\unit[303]{mm}). To extract the transmission $T$ of the monolayer, we measure the transmitted intensity through monolayer and substrate, and normalize it to the transmitted intensity of the bare substrate. To compare with the simulated absorption spectra, we calculate $1-T$ from our measurements, which yields the absorption of the monolayer, due to its very low reflectivity.

The absorption of the A and B excitons, as well as the higher excited states of the excitons overlap, depending on the material. To extract the line shape of the A excitons, all $1-T$ spectra are fit with curves for each resonance. The line shape of all resonances are broadened and asymmetric due to the phonon interaction. Therefore, we choose a Pearson IV distribution \cite{Kendall} to empirically describe the single resonances. 
\vspace{-0.5ex}
\begin{align}
 A(E) &= \frac{\left|\frac{\Gamma\!\left(m+\frac{\nu}{2}i\right)}{\Gamma(m)}\right|^2}{\alpha\, \mathrm{B}\! \left(m-\frac{1}{2}, \frac{1}{2}\right)}
\left[1 + \left(\frac{E-\lambda}{\alpha}\right)^{\!2\,} \right]^{-m}\times \nonumber \\
&\times\exp\left[-\nu \arctan\left(\frac{E-\lambda}{\alpha}\right)\right].
\end{align}

The first factor of the Pearson IV distribution is the normalizing constant, which involves the complex Gamma function $\mathrm{\Gamma}$ and the Beta function $\mathrm{B}$. The parameter $\lambda$ is the localisation parameter and represents the center of the peak, while $\alpha>0$ is a scale parameter, which defines the width of the function. The parameter $\nu$ describes the asymmetry or skewness of the function and $m>\frac{1}{2}$ the general shape. For $\nu=0$ the function is the symmetric Student's $t$ distribution. For $m=1$ the function is a skewed version of a Lorentzian, while for $m\to\infty$ the function becomes a skewed Gaussian. The scale and the shape parameters also influence the maximum of the distribution, which is given by $\lambda-\frac{\alpha\nu}{2m}$ \cite{Kendall}.

The extracted line shapes for all excitonic resonances are asymmetric, although for the B exciton and the excited states the asymmetry is barely visible, since these resonances are significantly broader than the A exciton resonance. The confidence of the fit depends mostly on the overlap between the excitonic resonances. Therefore, the line shape of monolayer WS$_2$, WSe$_2$, and MoSe$_2$ can be extracted with high confidence, while the fit for monolayer MoS$_2$ allows for some variation (FWHM \unit[59-69]{meV}). The best fit yields a FWHM of \unit[62]{meV}.

\begin{figure}[h]
\centering
\includegraphics[width=\textwidth]{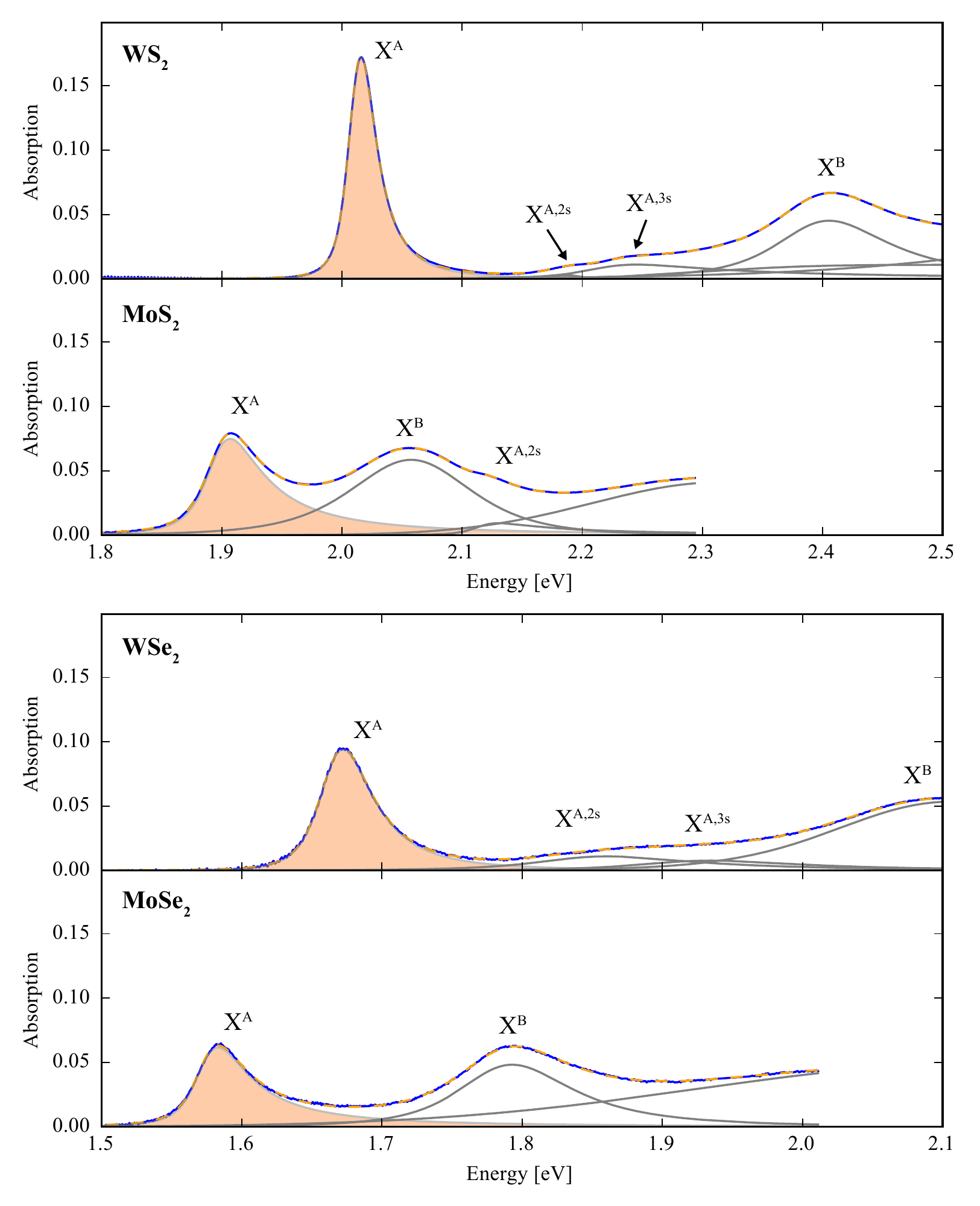}
\caption{Measured absorption spectra of monolayer WS$_2$, MoS$_2$, WSe$_2$ and MoSe$_2$ on borosilicate glass. The blue curve represents the experimental data, the dashed line the fit curve. Grey lines are the fit curves of the individial absorption resonances.}
\end{figure}

\section{Temperature-dependent optical transmission spectroscopy}

\begin{figure}[b]
\includegraphics[width=\textwidth]{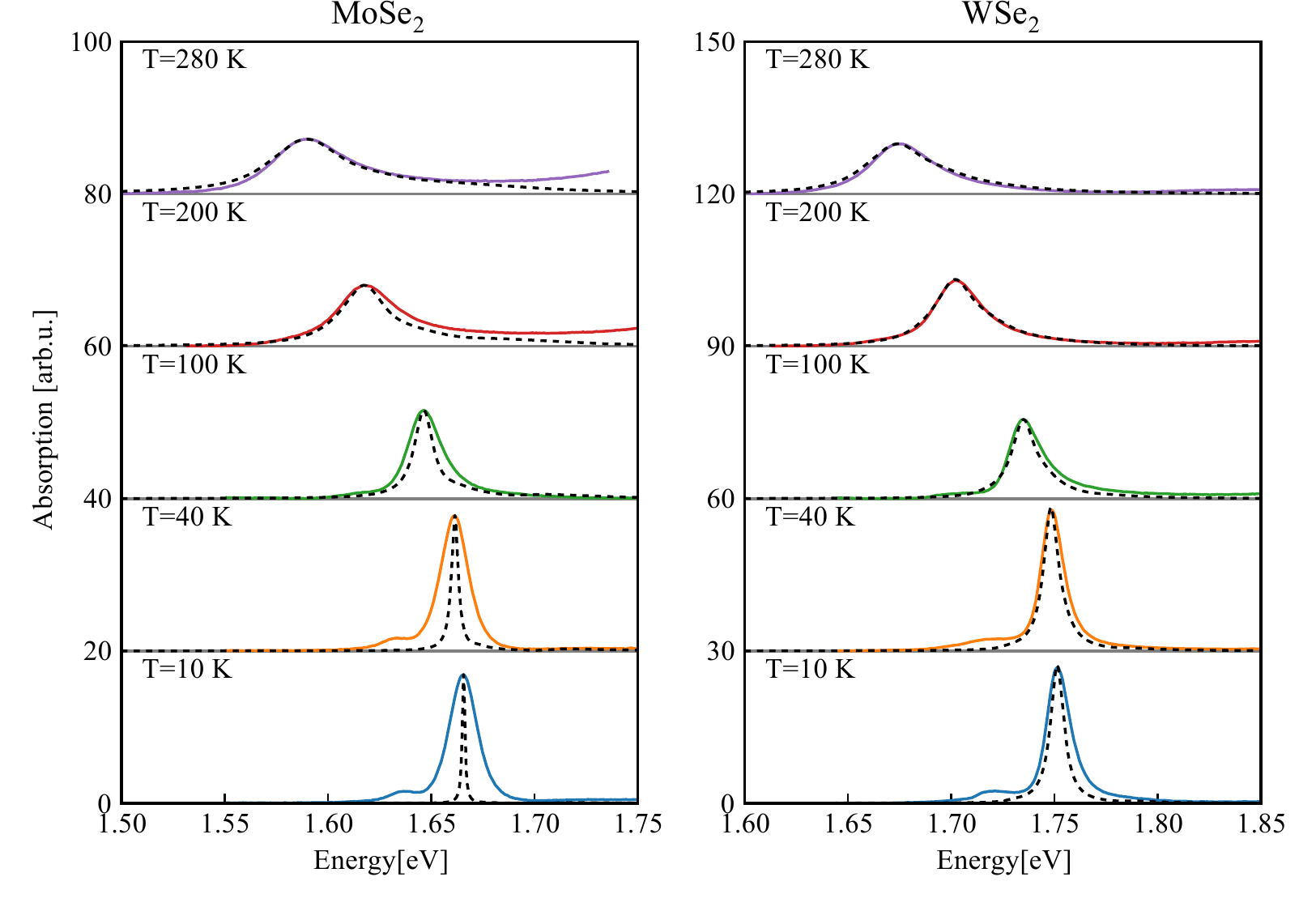}
\centering
\caption{Measured temperature-dependent absorption spectra of monolayer MoSe$_2$ and WSe$_2$ on sapphire substrate. The dashed line indicates the calculated curves, while the solid lines represent experimental data.
}
\label{temp}
\end{figure}

For temperature-dependent measurements (Fig. \ref{temp}) the samples are prepared by mechanical exfoliation from single crystals onto c-cut sapphire substrate with a thickness of \unit[500]{$\mu$m}. The transmission spectrum from the bare sapphire substrate, and the monolayer crystals on the substrate were measured for each temperature in a helium flow cryostat. On sapphire substrate the charged exciton (trion) is significantly stronger than on borosilicate glass due to substrate-induced doping. At temperatures of 200 K and above, the trion and exciton cannot be separated any more. Although the oscillator strength of the trion decreases with increasing temperature, there is still a finite contribution of the trion to the line shape at higher temperatures.

The measured spectra show stronger broadening than the simulated spectra at low temperatures due to disorder. With increasing temperature we see, based on the experimental spectra, that the asymmetry of the excitonic line shape is a temperature-dependent effect, emphasizing the phonon induced origin of the inhomogeneous broadening. The calculated spectra approach the experiment with increasing temperature until we find an excellent agreement at \unit[280]{K}.

As described in the main text, the calculated line shape of the A exciton of the MoS$_2$ monolayer is slightly narrower than the experimental data. We assume that this is caused by disorder-induced broadening. The MoS$_2$ monolayer is exfoliated from a natural crystal and is therefore expected to have a larger amount of impurities or crystal defects than the other monolayer materials. To include this disorder-induced inhomogeneous broadening we convoluted the theoretical curve with a Gaussian with $\sigma=10$ meV, represented by the dotted line. The width $\sigma$ of the Gaussian is chosen to match the FWHM of the theoretical curve with the experimental data. The convoluted curve is slightly shifted.

\begin{figure}[h]
\includegraphics[width=\textwidth]{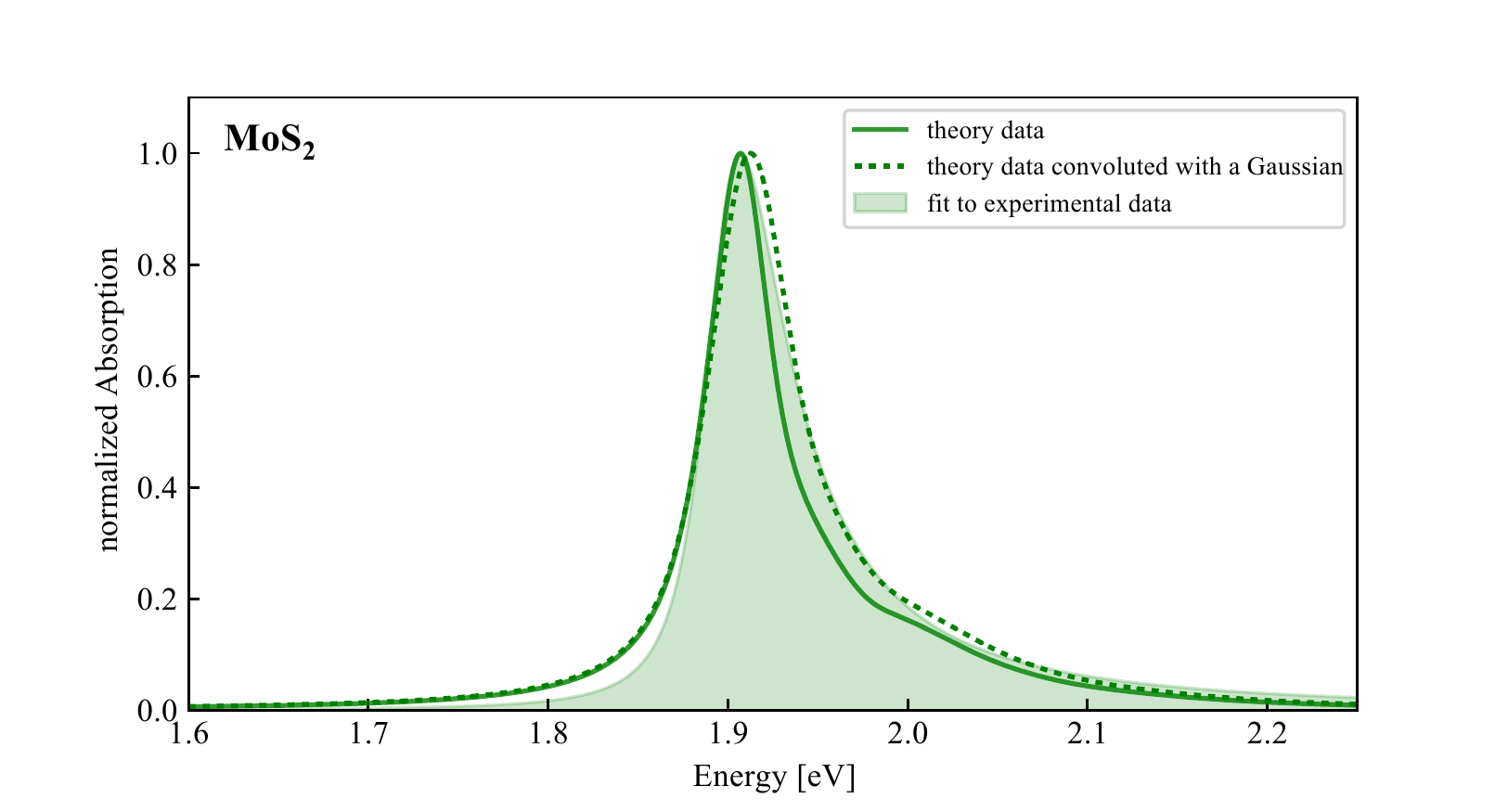}
\centering
\caption{Absorption spectra of monolayer MoS$_2$ on borosilicate glass. The filled curve represents the fit to the experimental data and the solid line the theoretical model. }
\label{convolute}
\end{figure}

\bibliographystyle{apsrev4-1}
%\bibliographystyle{abbrv}

%\bibliography{references}

%

\end{document}